\documentclass[conference]{IEEEtran}
\usepackage{cite}
\usepackage{amsmath,amssymb,amsfonts}
\usepackage{algorithmic}
\usepackage{graphicx}
\graphicspath{{figures/}}
\usepackage{textcomp}
\usepackage{xcolor}
\usepackage{soul}
\usepackage{geometry}
\def\BibTeX{{\rm B\kern-.05em{\sc i\kern-.025em b}\kern-.08em
    T\kern-.1667em\lower.7ex\hbox{E}\kern-.125emX}}
\newcommand*\mean[1]{\overline{#1}}

\makeatletter

\def\ps@IEEEtitlepagestyle{%
	\def\@oddfoot{\mycopyrightnotice}%
	\def\@evenfoot{}%
}
\def\mycopyrightnotice{%
	{\footnotesize DOI:10.1109/IJCNN.2019.8851887; 978-1-7281-2009-6/\$31.00~\copyright~2019 IEEE\hfill}
	\gdef\mycopyrightnotice{}
}

\begin{document}
\title{\vspace*{0.25in}Generative Adversarial Network for Radar Signal Generation\\
\thanks{Identify applicable funding agency here. If none, delete this.}
}

\author{\IEEEauthorblockN{Thomas Truong}
\IEEEauthorblockA{\textit{Department of Electrical and Computer Engineering} \\
\textit{University of Calgary}\\
\textit{Biometric Technologies Laboratory}\\
Calgary, Canada \\
thomas.truong@ucalgary.ca}
\and
\IEEEauthorblockN{Svetlana Yanushkevich}
\IEEEauthorblockA{\textit{Department of Electrical and Computer Engineering} \\
\textit{University of Calgary}\\
\textit{Biometric Technologies Laboratory}\\
Calgary, Canada \\
syanshk@ucalgary.ca}
}

\maketitle

\begin{abstract}
A major obstacle in radar based methods for concealed object detection on humans and seamless integration into security and access control system is the difficulty in collecting high quality radar signal data. Generative adversarial networks (GAN) have shown promise in data generation application in the fields of image and audio processing. As such, this paper proposes the design of a GAN for application in radar signal generation. Data collected using the Finite-Difference Time-Domain (FDTD) method on three concealed object classes (no object, large object, and small object) were used as training data to train a GAN to generate radar signal samples for each class. The proposed GAN generated radar signal data which was indistinguishable from the training data by qualitative human observers.
\end{abstract}

\begin{IEEEkeywords}
generative adversial networks, radar, concealed object detection, deep learning
\end{IEEEkeywords}

\section{Introduction}
Radar based methods are commonly used to non-destructively detect concealed objects. The application of radar based concealed object detection has been used in areas such as buried landmine detection \cite{Ko2012}, buried root detection \cite{Truong2018}, breast tumour detection \cite{Preece2016}, and concealed weapon detection on people \cite{Pitcher2018}. 

In security and access control systems, concealed object detection plays an integral part of ensuring public safety and security. There has been a recent trend in this field to use a multimodal screening procedure for deceptive behaviour \cite{Abouelenien2017}. Generally, these procedures involve computer vision problems which frequently apply machine learning methods to automate the process. The application of machine learning methods to automate radar based concealed object detection using has been limited by the lack of availability in high quality radar signal data.

Generative Adversarial Networks (GAN) have been a popular method of unsupervised learning in computer vision in recent years. Recent research on GANs have been focused on image generation and, as a result, GANs for one-dimensional data are still in the early stages of development. Moreover, there have been recent endeavors which analyze the utility GANs for data augmentation \cite{Antoniou2017}. Applying advanced machine learning algorithms such as deep neural networks requires a large amounts of data. In the absence of such data, which is currently the case on current radar based concealed object detection methods, these algorithms fail to exceed the performance of human inspection of radar data, which is labourious and expensive. As such, this paper presents the design a proof of concept for the use of GANs in radar signal generation with a focus on concealed object detection and seeks to establish the foundations for further research into the application of GANs for radar signal generation.

This paper is structured as follows: Section \ref{GANResearch} provides an overview of radar-based concealed object detection, GANs, and GAN applications for radar data. Section \ref{methodology} covers the design of the experiments to train the proposed GAN. Section \ref{results} summarizes the selected GAN model for radar signal generation. Sections \ref{conclusion} and \ref{future} summarizes the proposed GAN design and results and provides key research directions and questions to be answered in future works.

\section{Literature Review}\label{GANResearch}
\subsection{Radar Based Concealed Object Detection for Security and Access Control Systems}
Most current radar based algorithms for concealed object classification use simulated data which are free of clutter and generally only contain simple noise sources (such as additive white gaussian noise) \cite{Hutchinson2015, Lee2017}. As such, the majority of research on experimental radar signal classification problems for concealed object detection on humans are done in anechoic chambers and are focused on noise removal methods \cite{Hutchinson2015,Selver2016,Xia2013}. False positives from prostheses or medical implants such as pacemakers are also of concern for these systems. To scale these technologies for application in machine learning, particularly for training deep neural networks for classification, there is a need for high quality datasets which are tedious, time-consuming, and potentially expensive to collect. An unexplored application of a GAN is to generate one-dimensional radar signal data for data generation and augmentation.

\subsection{Generative Adversarial Networks} 
The original GAN formulation by Goodfellow et al.\cite{Goodfellow2014} consists of a discriminator network with input sample $x$ and output probability $D(x)$ and a generator network with input $z$ and output sample $G(z)$ with the same dimensions as $x$. The discriminator is trained to maximize the probability of labeling the generator samples $G(z)$ as \textit{fake} and the training data as \textit{real}. The generator is trained to maximize the probability that its samples $G(z)$ are labeled as \textit{real} by the discriminator. $D$ and $G$ are described to be playing a minimax game with the value function $V(G,D)$ which is defined by (\ref{minimaxValue}).

\begin{equation}
\small{\underset{G}{\text{min}}\ \underset{D}{\text{max}}V(D,G) = \mathbb{E}\big[\log{D(x)\big]} +  \mathbb{E}\big[\log{(1-D(G(z)))\big]}}
\label{minimaxValue}
\end{equation}

GANs have seen widespread application in the field of image processing and unsupervised image generation \cite{Radford2016,Reed2016}. Recently, GANs have seen application to produce one-dimensional data (similar to radar signal data) in audio applications \cite{Donahue2018, Anonymous2019}. Audio data, like the ultra-wideband radar signals used for object detection, are complicated, non-stationary signals which are prone to external sources of noise and are difficult to process. Moreover, the quality of the samples in these fields are difficult to quantitatively assess, often requiring qualitative human analysis to analyze the results. This gives rise to interest in developing GANs for radar signal data.

\subsection{Previous works: Generative Adversarial Networks for Radar Data}
Most applications of GANs and neural networks using radar data focus on images generated from radar signals using synthetic aperture radar (SAR) \cite{Marmanis2017,Guo2017,Schwegmann2017} and time-of-flight algorithms \cite{Brockner2018}. At the time of this paper, to the best of our knowledge, the application of GANs for one-dimensional radar signal data has not been published. 

\section{Methodology} \label{methodology}
The goal of our research is to provide a proof of concept for the application of GANs for radar signal generation. The proof of concept will use radar data collected from simulations to train a GAN to generate radar signals. A successful proof of concept will establish the foundations for further research on GAN applications in radar systems. In the future, we envision GANs to be a useful tool for data augmentation on radar data where data collection may be tedious, time-consuming, and expensive. Applications would include data augmentation on rare events such as buried explosive detection in the ground and concealed object detection on people.

\begin{figure}[ht!] 
	\centerline{\includegraphics{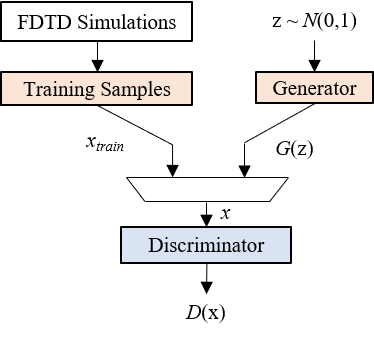}}
	\caption{Abstracted block diagram for the proposed GAN for radar signal generation.}
	\label{GANBlockDiagram}
\end{figure}

For our system, we modify the original GAN formulation to generate radar signals. Fig. \ref{GANBlockDiagram} shows the highly abstracted block diagram for the proposed GAN for radar signal generation. The training samples are focused on concealed object detection on humans and were generated using a Finite-Difference Time-Domain (FDTD) method are described in subsection \ref{simulationData}. The training samples contained 3 classes: no concealed object, large concealed object, and small concealed object. A separate GAN is trained for each class. Details on the structure, training, and evaluation methodology of the generator and discriminator blocks are described in subsection \ref{GANStructure}.

\subsection{Training Data Simulations} \label{simulationData}
The training data used to train the GAN is produced by implementing the FDTD method to numerically solve Maxwell's equations. The FDTD method is extremely popular in the field of computational electromagnetics and the details on implementing the FDTD method are covered extensively in Allen Taflove's "Computational Electrodynamics, the finite-difference time-domain method"\cite{A.Taflove}. As such, we will not be covering the FDTD method in this paper.

\begin{figure}[ht!] 
	\centerline{\includegraphics{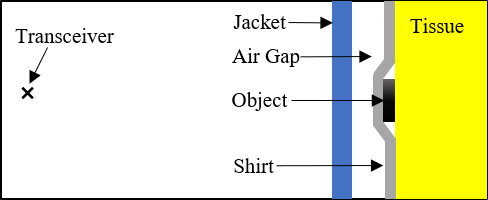}}
	\caption{An example of the system under test with a concealed object.}
	\label{FDTDSetup}
\end{figure}

For our training data we simulate the electrodynamics of an emitted 3.1 GHz - 5.3 GHz ultra-wideband pulse on a 2-dimensional system consisting of a jacket layer, an air gap layer, a shirt layer, a concealed object, and a tissue layer. Fig. \ref{FDTDSetup} shows an example of the described system under test with a concealed object. The shirt layer is flush against the tissue layer for samples with no object. Note that Fig. \ref{FDTDSetup} is not to scale and some features are enlarged for visibility. The system is designed to emulate a simplified real-life scenario where a suspect is attempting to conceal a highly reflective object underneath layers of clothing. The simulations are simulated over 20 cm on the vertical axis and 50 cm on the horizontal axis with absorbing boundary conditions. Table \ref{layerParams} contains the simulation parameters for the layers in system under test. 

\begin{table}[h!]
	\caption{Simulation Layer Parameters}
	\begin{center}
		\begin{tabular}{|c|c|c|}
			\hline
			\textbf{Layer}  & \textbf{Thickness (cm)}  & \textbf{Relative Permittivity (unitless)} \\ \hline
			Jacket & unif(1.5,2.5)   & 3                                \\ \hline
			Shirt  & unif(0.37,0.63) & 1.5                              \\ \hline
			Air Gap & unif(2.1,3.1)   & 1            \\   \hline
			Tissue & 10   & 40            \\   
			\hline               
		\end{tabular}
		\label{layerParams}
	\end{center}
\end{table}

Table \ref{objectParams} contains the concealed object parameters for the three major classes defined in our training data. Variation between separate samples are caused by the randomly generated thicknesses of the jacket, shirt, and air gap layers. For the large object and small object classes, additional variations between samples are caused by the randomly determined by the vertical object center position. For simplicity, the relative permittivity of the materials are set to be constant.

\begin{table}[h!]
	\caption{Simulation Object Parameters}
	\centering
	\begin{tabular}{|c|c|c|c|c|}
		\hline
\textbf{Class}                                          & \textbf{\begin{tabular}[c]{@{}c@{}}Thickness \\ (cm)\end{tabular}} & \textbf{\begin{tabular}[c]{@{}c@{}}Relative \\ Permittivity \\ (unitless)\end{tabular}} & \textbf{\begin{tabular}[c]{@{}c@{}}Object \\ Height \\ (cm)\end{tabular}} & \textbf{\begin{tabular}[c]{@{}c@{}}Object \\ Center \\ Position \\ (cm)\end{tabular}} \\ \hline
\begin{tabular}[c]{@{}c@{}}Large \\ Object\end{tabular} & 1                                                                  & 60                                                                                      & 5                                                                         & unif(4.5,16)                                                                          \\ \hline
\begin{tabular}[c]{@{}c@{}}Small \\ Object\end{tabular} & 1                                                                  & 60                                                                                      & 2.5                                                                         & unif(3.5,16.5)                                                                        \\ \hline

	\end{tabular}
	\label{objectParams}
\end{table}

\begin{figure}[hb!] 
	\centerline{\includegraphics[width = \linewidth, trim={0 1.4cm 0 1.0cm},clip]{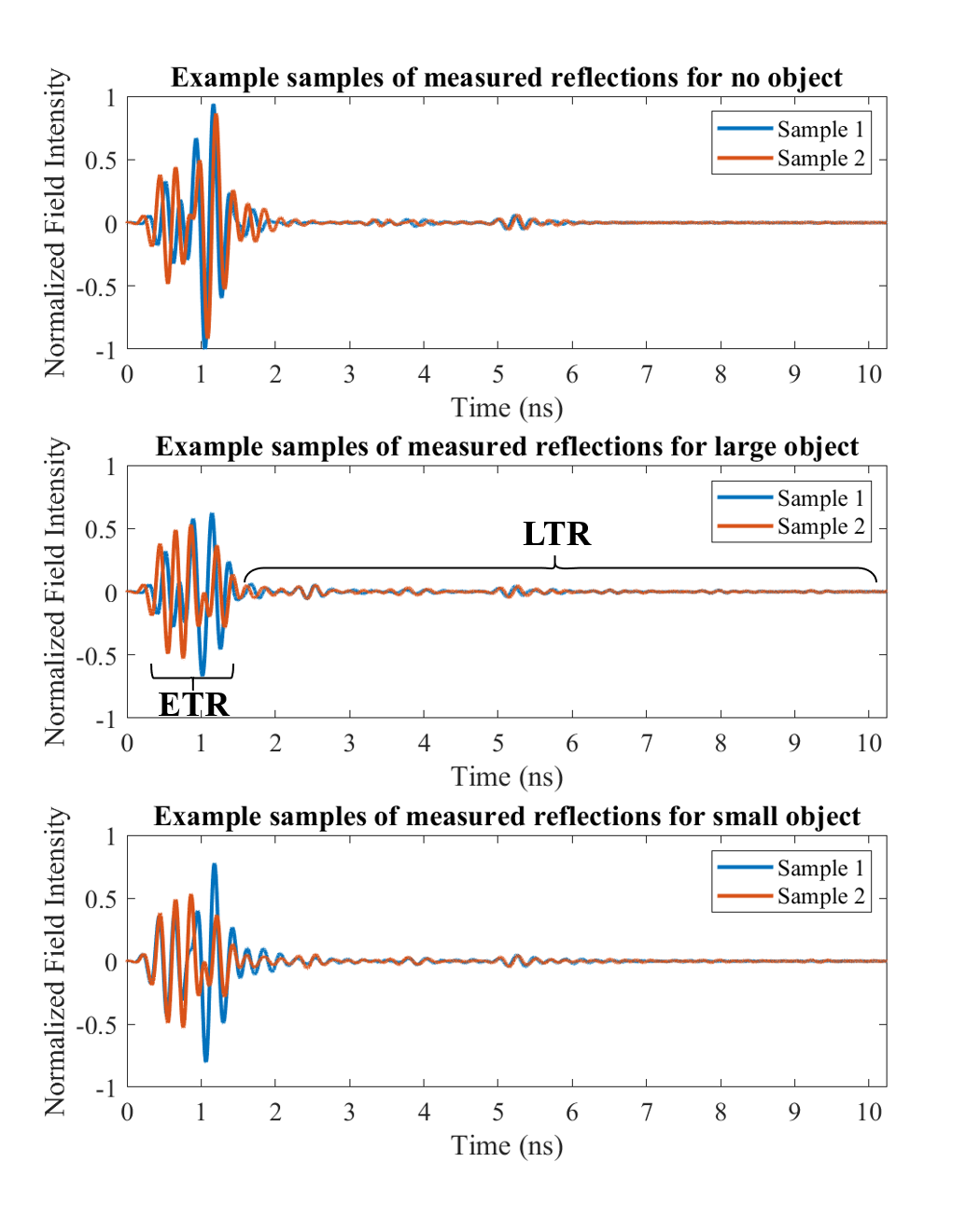}}
	\caption{Example training samples for the reflections measured at the transceiver for no object (top), large object (center), and small object(bottom) systems under test.}
	\label{trainingData}
\end{figure}

Fig. \ref{trainingData} shows two example samples for each of the classes simulated. Notice that the samples vary between samples of the same class given only small changes in layer thicknesses. These measured variations become exceedingly difficult to model without the use of computationally expensive electromagnetic simulation methods such as FDTD. The figure are annotated with the Early Time Response (ETR) and the Late Time Response (LTR). The ETR exists within approximately the first $1.5$ ns of the reflected signal and often captures the first reflections of the source signal off the system under test. The LTR consists of the the measured response after $1.5$ ns and contains smaller amounts of energy which have had multiple transmissions and reflections between layers in the system under test before returning to the transceiver. We simulated 3000 samples with a large object, 3000 samples with a small object, and 6000 samples with no object. These three classes of data are used as the training data used to train our GANs. The dataset is available upon email request.

\begin{figure}[ht!] 
	\centerline{\includegraphics[width = \linewidth, trim={0 1.4cm 0 1.0cm},clip]{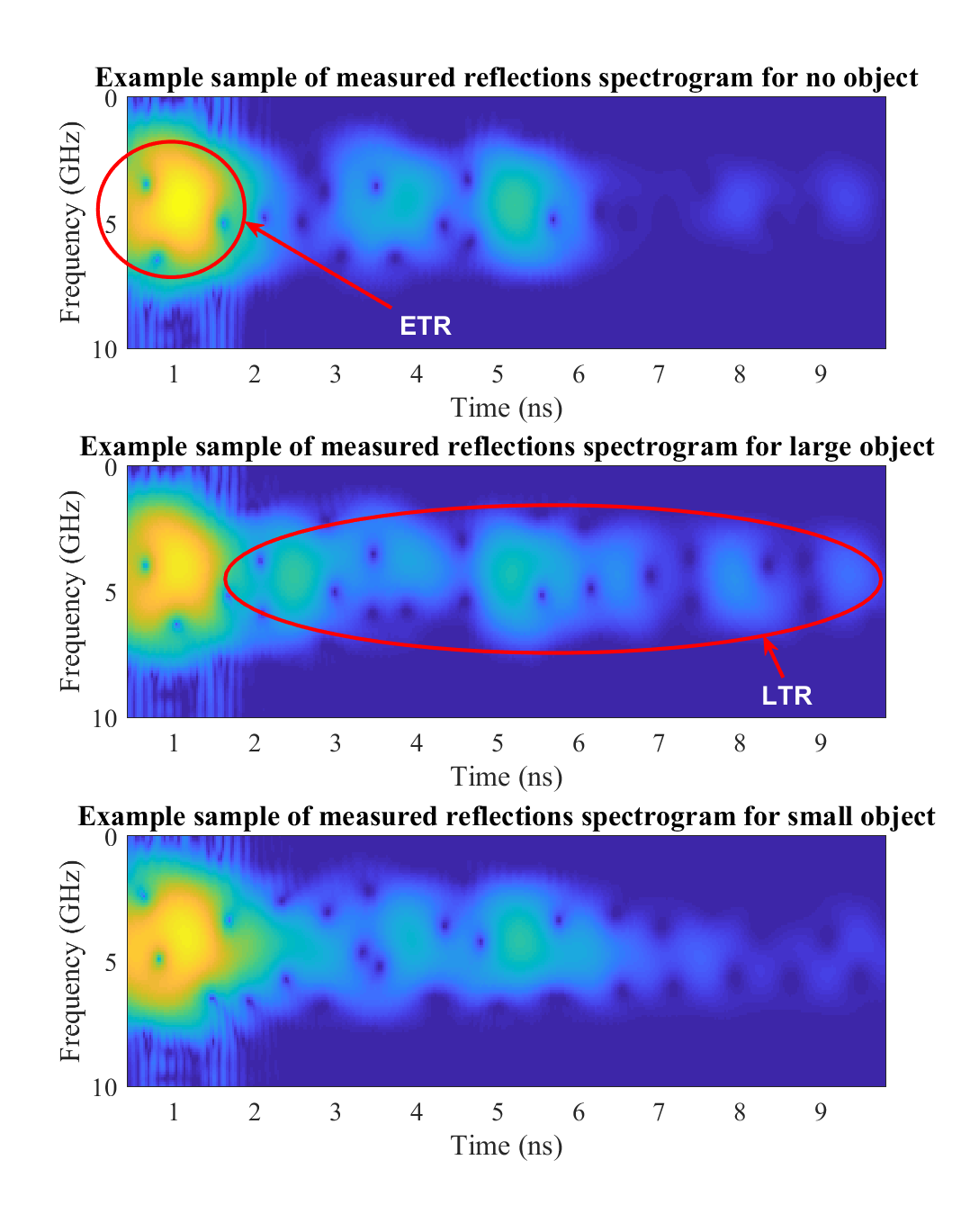}}
	\caption{Example spectrograms for each class computed using the simulated samples.}
	\label{simulatedSpecSamples}
\end{figure}

Fig. \ref{simulatedSpecSamples} shows sample spectrograms generated using MATLAB's \textit{spectrogram} function with a 700 time sample length window and 680 overlapped time samples \cite{MATLAB}. Spectrograms are a useful tool commonly used in audio analysis, and application here reveals visual differences between each of the classes of data simulated. The no object class contains little signal energy in the 3.1 GHz - 5.3 GHz frequencies past 6 ns. The large object class contains significant energy in those frequencies past 6 ns. The small object class contains energy around 4.0 GHz to 6.0 GHz past 6 ns.

\subsection{Generative Adversarial Network Structure and Training Methodology} \label{GANStructure}
\begin{figure*}[ht!] 
	\centerline{\includegraphics[width=\linewidth]{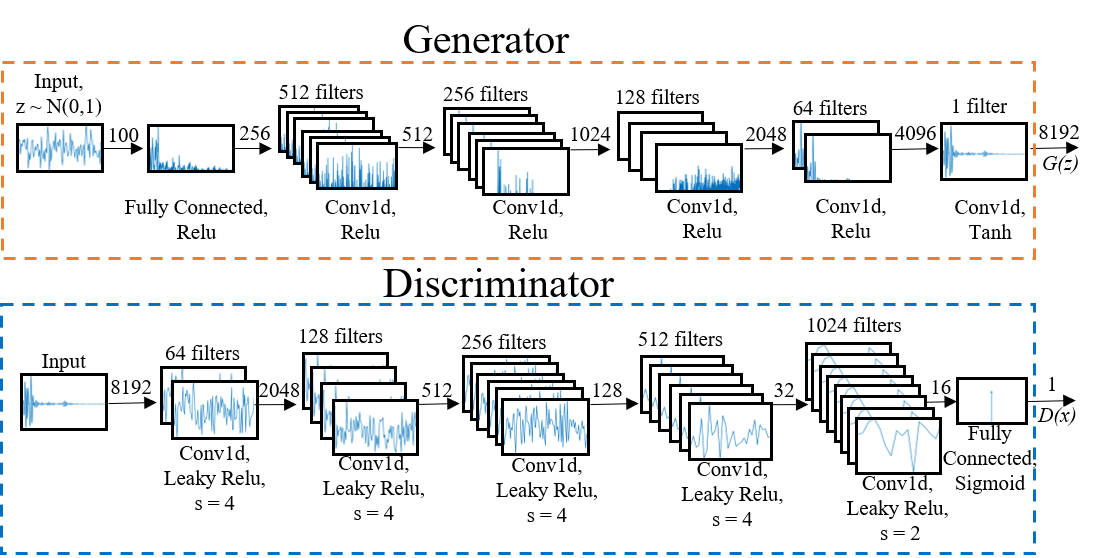}}
	\caption{Architecture for the proposed GAN.}
	\label{ganArchitecture}
\end{figure*}

Fig. \ref{ganArchitecture} shows the network architectures for the proposed GAN for radar signals. The architectures for the generator and discriminator are based off of WaveGAN \cite{Donahue2018} and DCGAN \cite{Radford2016}. For the generator, each convolutional layer ($s=1$ and $kernel\_size = 25$ for all convolutional layers in the generator) is preceded by an upsampling layer ($L=2$ by repeating each temporal step $2$ times for all upsampling layers) are used to output radar signals of length 8192. The discriminator contains a series of convolutional layers ($s = 4$ for all layers except for the final convolutional layer which has $s=2$) with kernel length 25 and stride length 4 to reduce the 8192 input sequence to a probability $D(x)$. Keras with a Tensorflow backend is used to implement these models in Python \cite{chollet2015keras}.

GANs in their original formulation are difficult to train as detailed in our literature review. The discriminator in our proposed GAN structure labelled samples with high confidence very early in training, resulting in insufficient gradient for the generator to learn. Ultimately, this caused the generator loss function to saturate almost immediately after the training starts. This produced poor results with no signs of improvement between each epoch. 

To overcome this early training problem we designed a two stage training process for our GAN. For stage one, we found that using a mean-squared error loss function helped prevent saturation of the generator loss function. Additionally, to help prevent an overly confident discriminator, we label the training samples $x_{train}$ as only 95\% likely to be real when training. In stage two, we qualitatively select the best generator model produced from stage one and reset the model on the discriminator using the default Keras layer initializers (\textit{glorot uniform}). We experimentally found that a binary cross-entropy loss function provided better results based on our qualitative and quantitative assessments. The hypothesis behind resetting the discriminator model is that the reset prevents overfitting of the discriminator on the training data. In the future, to further test this hypothesis and improve the training methodology, the use of dropout layers and regularization methods should improve the results, and may even remove the need for the two stage process.

Three GANs are trained, one for each class (no object, large object, small object) of data. Batch sizes of 30 are used for the large and small object classes and 60 is used for the no object class. Adam optimizer is used with hyperparameters $\alpha = 0.0001$, $\beta_1 = 0.5$, $\beta_2 = 0.9$, $\epsilon = 10^{-8}$ for both the discriminator and the generator. In early (stage one) training, we used $\alpha = 0.001$. For both stage one and stage two training, we train for approximately 400 epochs each, saving a model every 5 epochs. Training consists of periodically locking the generator layers and training the discriminator layers then locking the discriminator layers and training the generator layers. Training of the generator layers is done by passing in generator samples to the locked discriminator incorrectly labeled as \textit{real} and measuring the associated loss. In the future, the training process will be improved with the addition of stopping conditions and model quality to be discussed in Sections \ref{results} and \ref{conclusion}

The final generator models selected for each data class are based on human qualitative assessments and also a quantitative comparison of the statistical distributions of the training data and the generated signals for each epoch. Initially, we qualitatively select the models which generated samples which were mistaken by us in blinded tests to be training data samples. These qualitative assessments involved blinded tests of the raw waveform $x(t)$.

The ensemble variance is calculated using (\ref{ensembleVarianceEq}) for the selected generator models across $N = 3000$ generated samples. The ensemble mean, $\mean{x(t)} = \frac{1}{N}\sum_{i = 1}^{N}x_i(t)$ where $x_i(t)$ is sample number $i$, is used for these calculations.

\begin{equation}
\text{Var}[x(t)] = \mathbb{E}\Big[\big(x(t)-\mean{x(t)}\big)^2\Big] = \frac{1}{N}\sum_{i = 1}^{N}\big(x_i(t)- \mean{x(t)}\big)^2
\label{ensembleVarianceEq}
\end{equation}

Then, we use a similarity metric to compare the ensemble variances. We use the mean squared error (MSE), calculated using (\ref{MSE}), to compare the ensemble variances of the training samples and generated samples. In the future, the quality of other similarity metrics and also other statistical metrics should be analyzed.

\begin{equation}
\text{MSE} = \frac{1}{n} \sum_{t=1}^{n}\big(\text{Var}[G(z,t)]-\text{Var}[x_{train}(t)]\big)^2 
\label{MSE}
\end{equation}

Equation (\ref{MSE}) is computed with time gated signals resulting in $n = 8192$ time samples. The final generator model for each class is the generator model with the smallest mean squared error when compared to the training samples.

\section{Results} \label{results}
\begin{figure}[h!] 
	\centerline{\includegraphics[width = \linewidth, trim={0 1.4cm 0 1.0cm},clip]{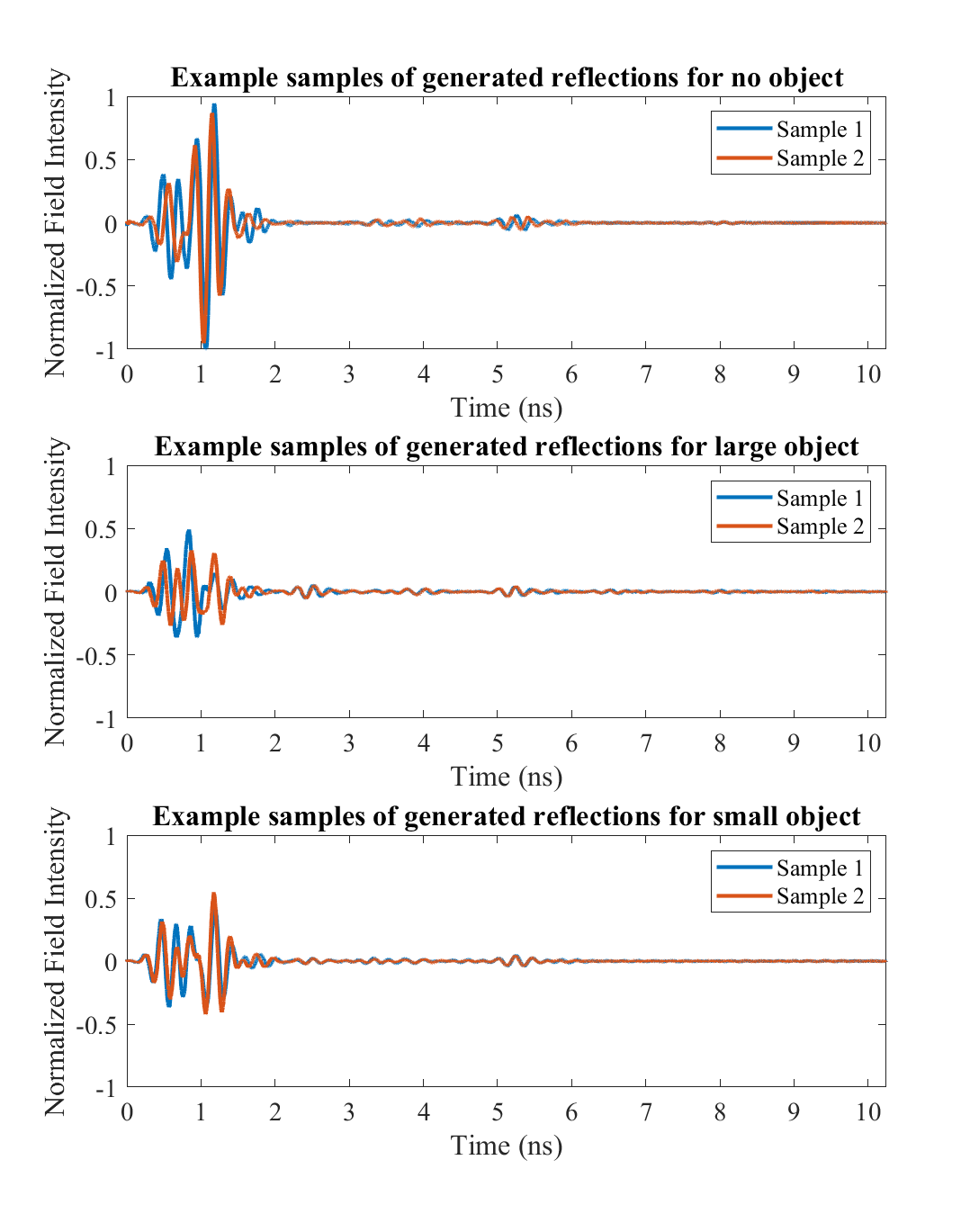}}
	\caption{Example generated samples for the reflections measured at the transceiver for no object (top), large object (center), and small object (bottom) systems under test.}
	\label{generatedSamples}
\end{figure}

\begin{figure}[h!] 
	\centerline{\includegraphics[width = \linewidth, trim={0 1.4cm 0 1.0cm},clip]{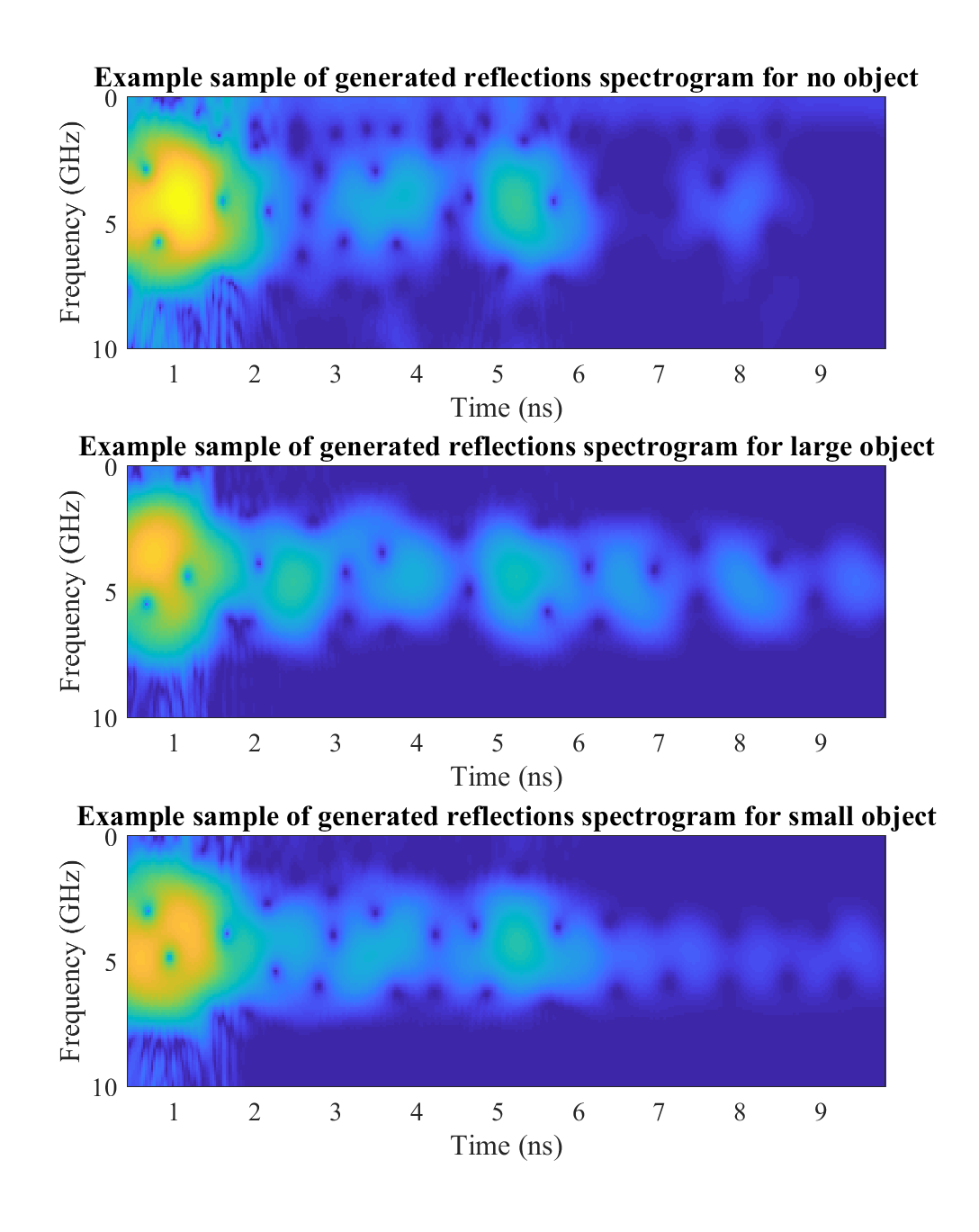}}
	\caption{Example spectrograms for each class computed using the generated samples.}
	\label{generatedSpecSamples}
\end{figure}

Fig. \ref{generatedSamples} shows two samples generated for each class and fig. \ref{generatedSpecSamples} shows sample spectrograms of the generated samples. Both figures use the samples generated by selected generator model. For the case of fig. \ref{generatedSpecSamples}, the spectrogram was determined from Sample 1 of each of the classes. The saved Keras generator models used to create these samples are available by email request. At a qualitative glance, the GAN performed well in capturing the distribution of the training data. A qualitative comparison of Figs. \ref{trainingData} and \ref{simulatedSpecSamples} from the training data and Figs. \ref{generatedSamples} and \ref{generatedSpecSamples} from the generated data show no features in the generated samples which would allow a human being to differentiate between the two.

\begin{figure}[h!] 
	\centerline{\includegraphics[width = \linewidth, trim={0 1.4cm 0 1.0cm},clip]{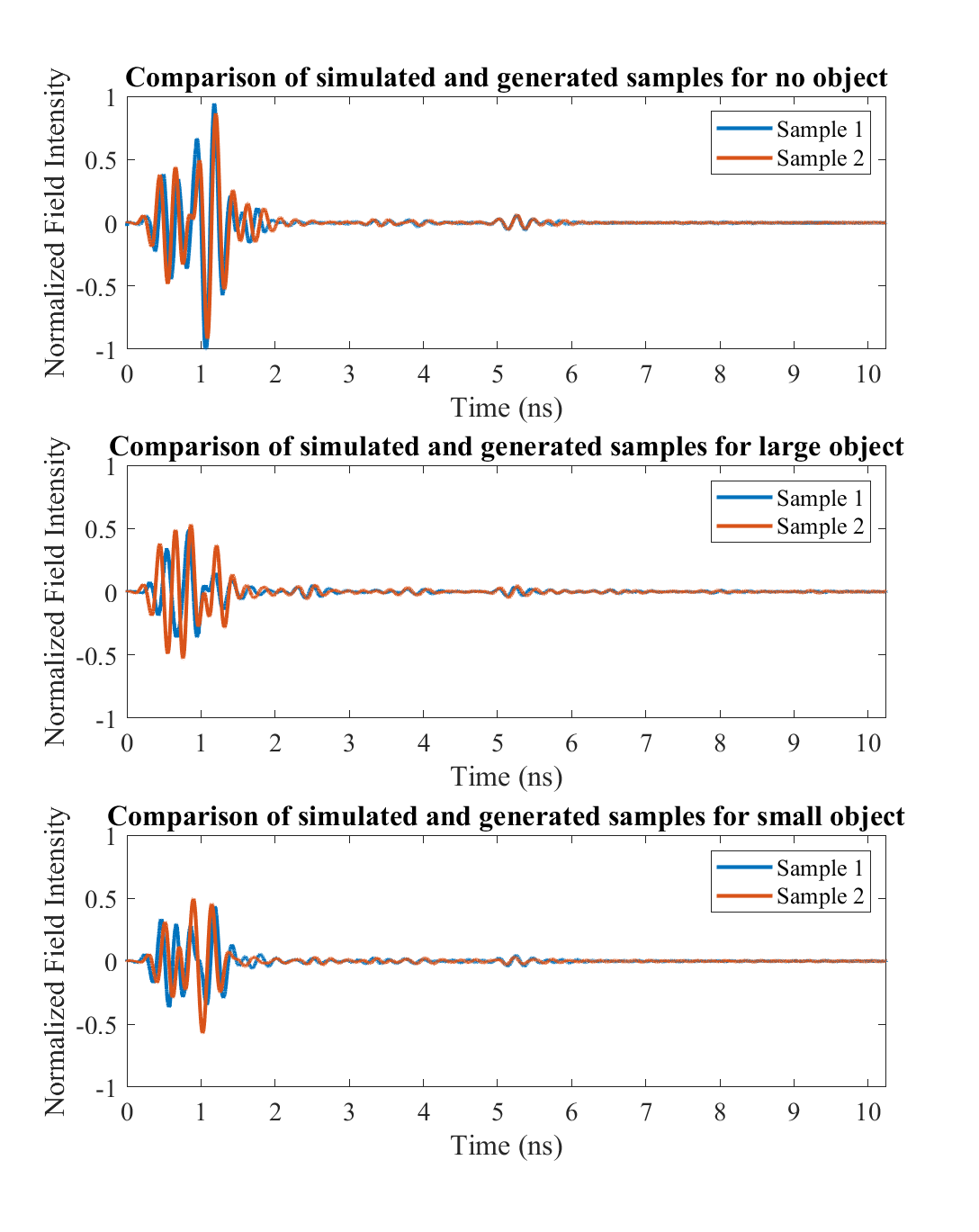}}
	\caption{Comparison of training samples and generated samples for the reflections measured at the transceiver for no object (top), large object (center), and small object (bottom) systems under test. An exercise for the reader: For each of the classes, identify which sample is from the training data and which is generated by the GAN. Answer is in Section \ref{results}.}
	\label{comparisonSamples}
\end{figure}

Fig. \ref{comparisonSamples} shows a comparison of the training samples and is an example of the blind test used for model selection. In Fig. \ref{comparisonSamples}, for all three classes, Sample 1 is the generated sample by the GAN and Sample 2 is the training sample.

\begin{figure}[h!] 
	\centerline{\includegraphics[width = \linewidth, trim={0 1.4cm 0 1.0cm},clip]{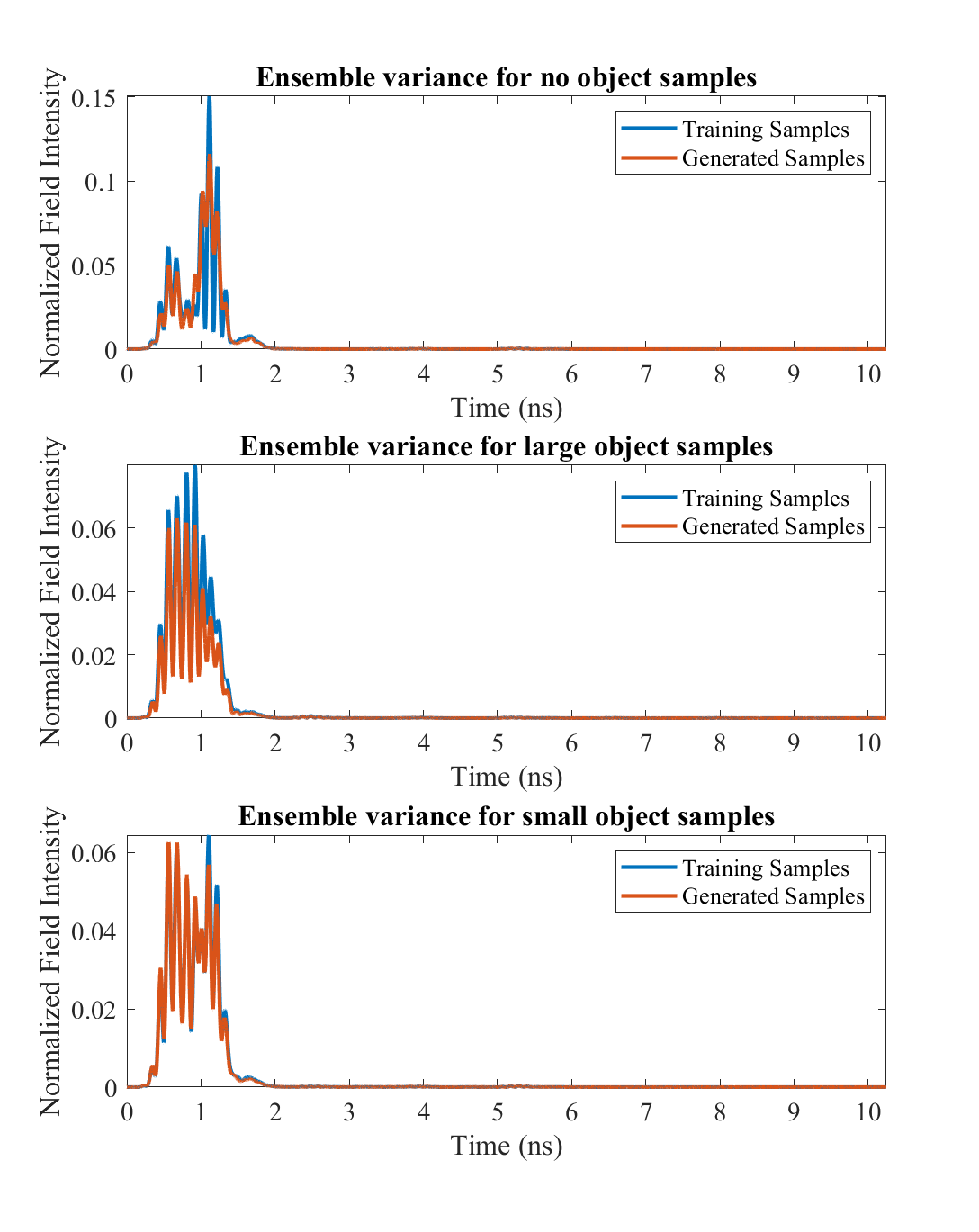}}
	\caption{Comparison of the ensemble variances between the training and generated samples for no object (top), large object (center), and small object (bottom) systems under test.}
	\label{ensembleVariance}
\end{figure}

Fig. \ref{ensembleVariance} shows the ensemble variances of the GAN model calculated using (\ref{ensembleVarianceEq}) from Section \ref{GANStructure}. The calculated MSE is $3.3\mathrm{e}{-5}$ for the no object generator, $1.2\mathrm{e}{-5}$ for the large object generator, and $9.0\mathrm{e}{-7}$ for the small object generator. 

\section{Discussions and Conclusions} \label{conclusion}
This paper presented a proof of concept for the application of GANs for radar signal generation. To the best of our knowledge, this is the first application of a GAN to the field of radar signal generation. Initially, training data is simulated  using the FDTD method to model a system under test which emulated a simplified real-life scenario on three radar signal classes: one class containing no concealed object, one class containing a large concealed object, and one class containing a small concealed object. This training data is used to train a GAN which attempts to generate samples that replicate the training data distribution. The results of the GAN show promising results for the generation radar signal data, generating samples which are indistinguishable (by humans) from the training samples. 

Moving forward, this proof of concept lays the foundation for future research into the field of radar signal generation using GANs. With additional research, GANs may be capable of performing data augmentation on tedious, time-consuming, and expensive to collect radar signals.

\section{Future Work} \label{future}
At this stage in our works, this paper has shown that a GAN is capable of replicating the distribution of radar signals to levels which are indistinguishable from the training data by a human; however, our methodologies are still primitive when compared to the advanced methodologies that have been developed for the application of GANs in other fields such as image generation. In this section we identify several key directions of research and questions to be answered which will dictate the success of the application of GANs in radar signal generation.

A potentially major obstacle for the application of GANs in radar signal generation is its ability to generate signals given noisy experimentally collected radar data. The simulated data from this paper are simulated with very ideal noise-free radar conditions. Given noisy training data with a poor signal-to-noise ratio, is it possible for a GAN to capture the distribution of the training data? In addition to a GANs applicability to noisy and low quality training data, how well can a GAN be trained to recognized a training data's distribution given limited radar training data? Will common methods such as transfer learning be a potential option for a GAN when dealing with limited training data?

Beyond a GANs applicability on experimental training data, how can the training process of these GANs be improved for the purposes of radar signal generation? Other GANs on one-dimensional data, particularly audio generation, rely heavily on qualitative measurements in analyzing their GAN outputs \cite{Donahue2018,Anonymous2019}. For the case of radar signal generation, are statistical metrics and similarity metrics sufficient to be included in the training process to help determine stopping criteria? Will including these metrics in the training process improve the results on more difficult training data? Is the discriminator overfitting the training data and will implementation of dropout layers and application of regularization improve the training process? 

An interesting application of radar signal generating GANs is to try to produce reflected radar signals given the electrical and physical system under test parameters. In image generation, text to image synthesis has been done using a GAN by encoding input parameters into the generator input $z$ \cite{Reed2016}. Is it possible to encode system under test data, such as the layer thicknesses and electrical parameters from this paper in a similar fashion for radar signal generation? Will the GAN be able to extrapolate radar signals that don't belong to in the training data set that are accurate to the FDTD simulated results? Extending this idea, can the GAN produce experimental radar data given a colour image of the system under test?

\section*{Acknowledgment}

This work was partially supported by the Natural Sciences and Engineering Research Council of Canada through the Engage Grant and Discovery Grant "Biometric Intelligent Interfaces".

\bibliographystyle{IEEEtran}
\bibliography{IEEEabrv,references}

\end{document}